\documentclass{article}

\usepackage[preprint]{neurips_2025}

\usepackage[utf8]{inputenc} 
\usepackage[T1]{fontenc}    
\usepackage{hyperref}       
\usepackage{url}            
\usepackage{booktabs}       
\usepackage{amsfonts}       
\usepackage{nicefrac}       
\usepackage{microtype}      
\usepackage{xcolor}         
\usepackage{amsmath}
\usepackage{graphicx}
\usepackage{multirow}
\usepackage{multicol}
\usepackage{algorithm}
\usepackage{algpseudocode}
\usepackage{amsmath,amssymb}
\usepackage{makecell}

\newcommand*\samethanks[1][\value{footnote}]{\footnotemark[#1]}

\newcommand\clearrow{\global\let\rowmac\relax}
\clearrow

\title{A Graph Completion Method that Jointly Predicts Geometry and 
Topology Enables Effective Molecule Assembly}

\author{%
  Rohan V. Koodli\thanks{Address correspondence to: \texttt{koodli@stanford.edu}, \texttt{rondror@cs.stanford.edu}} \\
  Stanford University \\
  \And
  Alexander S. Powers \\
  Stanford University \\
  \AND
  Ayush Pandit \\
  Stanford University \\
  \And
  Chiho Im \\
  Stanford University \\
  \And
  Ron O. Dror\samethanks \\
  Stanford University \\
}

\begin{document}

\maketitle

\begin{abstract}
A common starting point for drug design is to find small chemical groups or ``fragments'' that form interactions with distinct subregions in a protein binding pocket. The subsequent challenge is to assemble these fragments into a molecule that has high affinity to the protein, by adding chemical bonds between atoms in different fragments. This ``molecule assembly'' task is particularly challenging because, initially, fragment positions are known only approximately. Prior methods for spatial graph completion—adding missing edges to a graph whose nodes have associated spatial coordinates—either treat node positions as fixed or adjust node positions before predicting edges. The fact that these methods treat geometry and topology prediction separately limits their ability to reconcile noisy geometries and plausible connectivities. To address this limitation, we introduce EdGr, a spatial graph diffusion model that reasons jointly over geometry and topology of molecules to simultaneously predict fragment positions and inter-fragment bonds. Importantly, predicted edge likelihoods directly influence node position updates during the diffusion denoising process, allowing connectivity cues to guide spatial movements, and vice versa. EdGr substantially outperforms previous methods on the molecule assembly task and maintains robust performance as noise levels increase. Beyond drug discovery, our approach of explicitly coupling geometry and topology prediction is broadly applicable to spatial graph completion problems, such as neural circuit reconstruction, 3D scene understanding, and sensor network design.

\end{abstract}

\section{Introduction}

In order to design a drug that targets a given protein, one must find molecules that bind tightly and specifically to this protein while maintaining properties such as synthesizability and solubility. This remains a challenging problem, even for target proteins whose structures have been available for decades.

One promising approach is to first find small chemical groups, known as ``fragments,’’ that interact favorably with various parts of a target protein binding pocket, and then assemble these fragments into a larger molecule that binds tightly to the target by adding chemical (covalent) bonds between them. Multiple methods are available to find such fragments, including experimental screening, intuitive design by medicinal chemists, and generative AI techniques \citep{lamoree2017current, sheng2013fragment, carloni2025rational, shim2011computational, medsage, neeser2025flowbased}. 

However, assembling these fragments into a larger molecule with high affinity for the target remains difficult. In machine learning terms, this is a graph completion problem, where the nodes are atoms, and the edges are bonds. One knows the edges within each fragment and must predict the best edges to connect the fragments to obtain a molecule that binds tightly and specifically to the target.

Given the precise spatial coordinates of atoms in a molecule, determining the missing bonds is relatively straightforward, as atoms connected by covalent bonds have characteristic distances, and pairs of bonds to a shared atom form characteristic angles. These characteristic bonds and angles can be determined easily given the types of atoms involved. 

Molecule assembly is difficult in practice because prior to assembling the fragments, one does not know the exact position or orientation of each fragment, leading to substantial uncertainty in atom positions. Experimental and computational methods for selecting fragments generally give only approximate information about fragment positions, and fragment positions and orientations may change substantially when multiple fragments are connected into a larger molecule.

A substantial body of previous work focuses on knowledge graph completion---predicting missing edges in a graph. These methods are widely used for purposes including recommender systems and analysis of social networks \citep{Zamini_2022}. Unfortunately, these methods either do not make use of spatial coordinates or assume that the spatial coordinates will not change as edges are added \citep{li2023evaluating, mao2023revisiting}.

Recent work on molecular design has included the development of denoising diffusion models for generating atom positions in a candidate drug \citep{schneuing2024structure, du2024machine, morehead2024geometry, guan2024decompdiff, schneuing2025multidomain}. These methods either add bonds after the diffusion process is complete or generate bonds via a separate diffusion branch, leading to no or weak coupling between the node positioning and bond predictions. A skilled medicinal chemist, on the other hand, would consider possibilities for inter-fragment bond formation while simultaneously attempting to position fragments relative to one another.

We thus developed EdGr, a spatial graph diffusion model that learns to simultaneously predict inter-fragment bonds and atomic coordinates to solve the molecule assembly problem. At every denoising step, EdGr directly predicts weights for each potential inter-fragment edge, which guide atoms toward positions favored by high-confidence edges. In turn, these updated positions are used to refine candidate edge weights.

EdGr substantially outperforms prior methods for molecule assembly, including both classical knowledge graph edge prediction methods \citep{li2023evaluating, mao2023revisiting} and recent spatial graph methods \citep{liao2022equiformer, hoogeboom2022equivariantdiffusionmoleculegeneration} that move atoms and use post-hoc methods \citep{landrum2013rdkit, o2011open} to add bonds once atomic coordinates have been set. EdGr works with a wide variety of chemical fragments, including individual atoms. EdGr can be extended to spatial graph completion problems where no initial coordinate estimates are provided, by simply placing the fragments or nodes at a pre-defined location and using an EdGr model trained with a very high level of coordinate noise.

\begin{figure}
    \centering
    \includegraphics[width=1.0\linewidth]{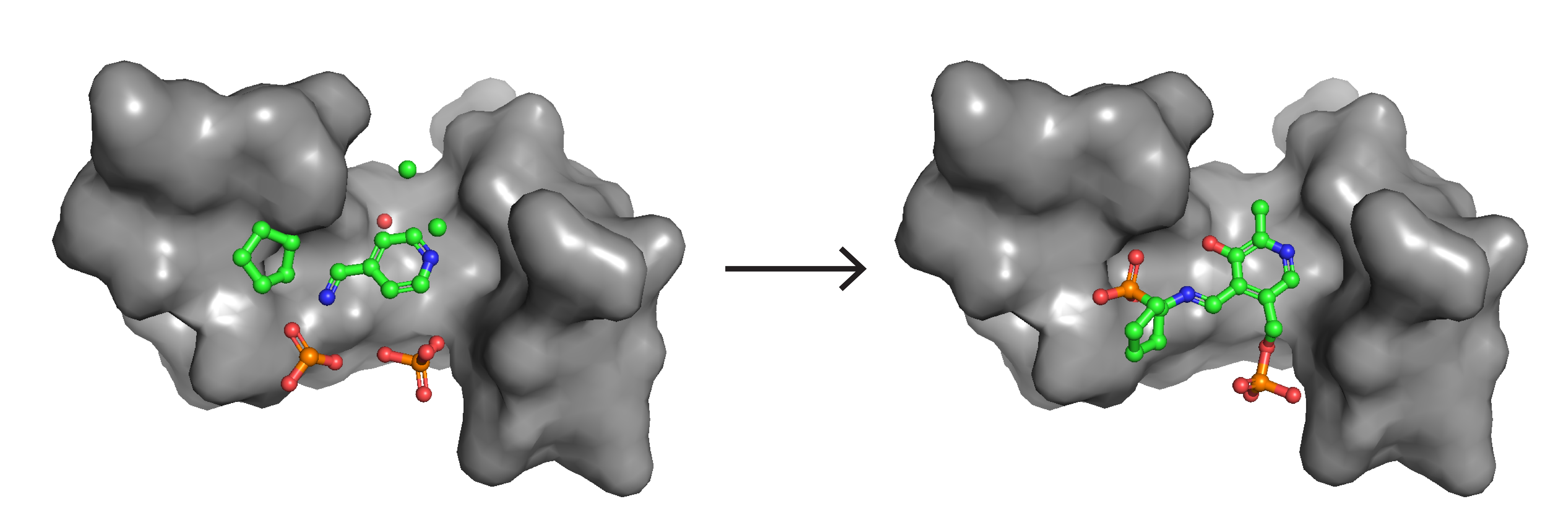}
    \caption{Molecule Assembly Problem Definition. Given fragments (small chemical groups) scattered in a protein binding pocket (left), we wish to predict inter-fragment chemical bonds that will connect them into larger molecule that binds tightly to the pocket (right). The fragments positions and orientations in the larger molecule differ from those provided initially.}
    \label{fig:molecule_assembly}
\end{figure}

\section{Related Work}

\subsection{Differences from fragment linking and all-atom molecule generation}

The molecule assembly task addressed in this paper is substantially different from two equally important drug design tasks that have previously attracted more attention in the machine learning community: fragment linking and all-atom molecule generation.

\paragraph{Fragment Linking} The fragment linking problem is defined as follows: given two fragments positioned precisely in a protein pocket, create a chain of atoms to link the fragments together. Multiple computational approaches have been developed for fragment linking, including database search \citep{sheng2013fragment}, autoregressive modeling \citep{imrie2020deep}, variational autoencoders \citep{huang20223dlinker}, and diffusion models \citep{igashov2024equivariant}. Fragment linking involves adding atoms between statically positioned fragments, where molecule assembly involves adding chemical bonds between fragments  whose positions can change.

\paragraph{All-atom molecule generation}

All-atom molecule generation is defined as follows: given a protein pocket, generate a high-affinity ligand, atom by atom. A variety of diffusion and flow-based models address this task by sampling atoms positions, atom identities (element types) \citep{schneuing2024structure, guan20233d}, and sometimes bond types \citep{guan2024decompdiff, morehead2024geometry}. The primary goal in all-atom molecule generation is to predict atom identities and coordinates, whereas the primary goal in molecule assembly is to predict bonds between fragments. Even in all-atom molecule generation methods that include a bond generation branch, it is weakly coupled to the atomic positions; the primary objective of these methods is to generate atoms that form a ligand, and making bonds is a side task. In molecule assembly, the opposite holds true: the primary objective is to generate new bonds, and fragment positions may be updated to facilitate these bond predictions.

\subsection{Existing Spatial Graph Completion Approaches}

Two classes of previous developed methods are available for spatial graph completion: topology-prediction methods and geometry-prediction methods.

\paragraph{Topology-prediction Methods} These methods predict missing edges in a graph, treating spatial coordinates as fixed. These methods can be further subdivided into two subclasses: ML-based link prediction methods and heuristic methods.

ML-based link prediction methods \citep{kipf2016semi, velivckovic2017graph} are commonly used for graph completion in the context of knowledge graphs \citep{Zamini_2022, chaudhri2021introduction}. These methods typically use graph neural networks to learn to impute missing edges in incomplete graphs.

Heuristic methods predict edges without any learnable parameters, simply relying on properties of the graph to make predictions. For example, the Common Neighbors \citep{newman2001clustering} heuristic computes the similarity of pairs of nodes and links nodes with the highest similarities, and the Minimum Distance heuristic connects nodes that are the closest in physical space.

\paragraph{Geometry-prediction Methods} \label{geompred} Geometry-prediction methods explicitly predict spatial coordinates of nodes; edges can then be inferred based on methods such as Minimum Distance. Geometry-prediction methods such as EDM \citep{hoogeboom2022equivariantdiffusionmoleculegeneration} and Equiformer \citep{liao2022equiformer} are explicitly designed for tasks in n-dimensional space, and are popular for molecular applications. The goal of these methods is to predict point positions and attributes, and they are able to do so by treating the points in space as nodes in a graph, with edges inferred via a distance cutoff (alternatively, the user can pre-specify edges if known beforehand).

\subsection{Diffusion Models}
Denoising Diffusion Probabilistic Models (Diffusion Models, or DDPMs) \citep{sohl2015deep, ho2020denoisingdiffusionprobabilisticmodels}, are generative machine learning models inspired by non-equilibrium thermodynamics. They are characterized by two processes: a \textit{forward noising} process which gradually adds Gaussian noise to the original data $x$ via a Markov chain; and a \textit{denoising} process which is parametrized by a neural network $\phi$ that learns to remove the noise.

\paragraph{Self Conditioning} In diffusion, $\phi$ learns to either remove the noise $\epsilon$ or directly predict $x_0$ in the chain of denoising steps. However, any intermediate predictions $\Tilde{x}_0$ are discarded in the subsequent diffusion steps; self conditioning addresses this deficiency \citep{chen2023analogbitsgeneratingdiscrete}. Instead of ignoring these intermediate predictions, self conditioning takes these predictions and concatenates them to the noise at timestep $t$ to provide additional context for the model, yielding much better downstream performance \citep{chen2023analogbitsgeneratingdiscrete}. To prevent the model from becoming too reliant on the intermediate $\Tilde{x}$s, we introduce stochasticity with a random variable $s \sim \textit{U}(0,1)$; if $s$ is greater than or equal to a preset threshold $p$, self conditioning is not applied.

\section{Methods}

\subsection{Forward Noising Process}
We note some key differences between EdGr's forward noising process and that used in standard spatial graph diffusion model. Unlike the standard case, where every point is noised following a closed form multivariate Gaussian distribution, our model treats fragments as rigid (because our fragments are basically rigid, all our fragments have 0 rotatable bonds); every atom in a fragment is noised according to the same translation and rotation vector.

Fragment translational noise is sampled the same way as a standard spatial graph diffusion model samples atom coordinate noise:

\begin{align}
    \label{eq:fwd_trans} q(z_t | x) &= \mathcal{N}(z_t | \alpha_t x, \sigma^2_t I)
\end{align}

$\alpha \in \mathbb{R}^+$ controls the amount of signal retained in original coordinates $x$ and $\sigma^2 \in \mathbb{R}^+$ controls the variance of the normal distribution, in Ångstroms. We do not add noise to the atomic features $h$ as the fragment and atom identities are fixed in molecule assembly. For noising a fragment's orientation, we follow the isotropic Gaussian distribution on SO(3) $g \sim \mathcal{IG}_{SO(3)}(\mu=0, \epsilon^2)$ \citep{leach2022denoising, savjolova}, which has the density function:

\begin{align}
    \label{eq:rot} f(\omega) &= \frac{1-\cos \omega}{\pi} \sum_{l=0}^{\infty} (2l+1)e^{-l(l+1)\epsilon^2} \frac{\sin ((l + \frac{1}{2})\omega)}{\sin (\frac{\omega}{2})}
\end{align}

\subsection{Model and Training Details}

\paragraph{Preliminaries} We define the following: $h^l$ are node embeddings at layer $l$; $x^l$ are node coordinate embeddings at layer $l$; $s$ is an indicator variable representing self conditioning; $a$ are predefined edge features; $\phi$ are neural networks; $m$ are known edge embeddings; and $n$ are missing edge embeddings. We include pocket atoms in our graph representation, but treat these atoms as static.

\paragraph{EdGr Implementation}

\begin{figure}
    \centering
        \includegraphics[width=1.0\linewidth]{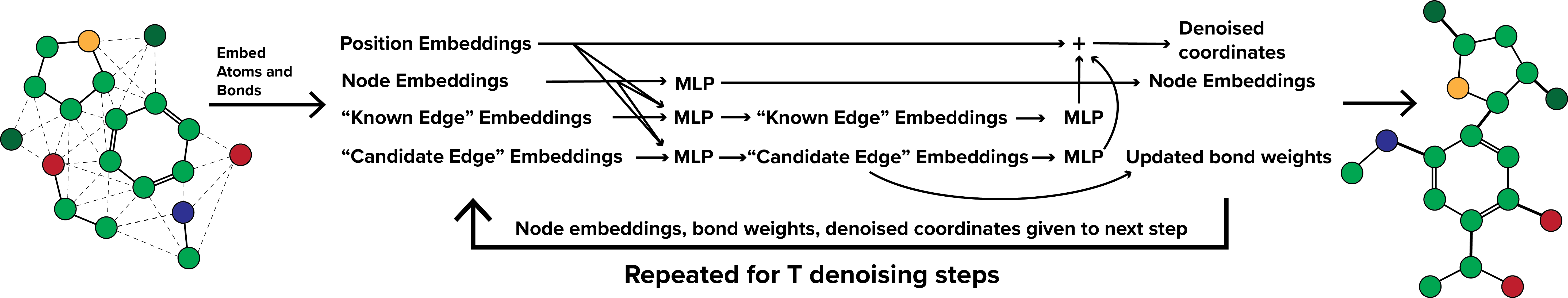}
    \caption{EdGr Architecture Schematic. Node information and edge information are learned through MLPs, and the model outputs updated positions and edge weights (middle). After repeating for T denoising steps, the full molecule with final positions and connectivity is produced (right).}
    \label{fig:model_architecture}
\end{figure}

A diagram of the EdGr model can be seen in Figure \ref{fig:model_architecture}; we take inspiration from the EGCL layers from EGNNs \citep{satorras2022enequivariantgraphneural}. We have two parallel multi-layer perceptrons (MLPs) to learn edge features, one for known edges and one for missing edges. The known and missing edge features then get aggregated per node and get passed to node MLPs that update node embeddings and positions.

\begin{align}
\label{eq:self_cond} s &= \begin{cases}
    1 \text{ if } \textit{U}(0, 1) < p \\
    0 \text{ otherwise}
\end{cases} \\
\label{eq:egcl}    m_{ij} &= \phi_e (h^l_i, h^l_j, ||x^l_i-x^l_j||^2,  a_{ij}) \\ 
\label{eq:e_if}    n_{ij;t} &= \phi_f (h^l_i, h^l_j, ||x^l_i-x^l_j||^2,  a_{ij}, s * n_{ij;t-1}) \\
\label{eq:egcl_last}     x^{l+1}_{i} &= x^{l}_{i} + \frac{1}{M-1} \sum_{j \ne i} (x^l_i - x^l_j) (\phi_x(m_{ij}) + \phi_y(n_{ij})) \\
\label{eq:egcl_agg} m_{i} &= \sum_{j \in \mathcal{N}(i)} m_{ij} \\
\label{eq:eif_agg}     n_{i} &= \sum_{j \in \mathcal{N}(i)} n_{ij} \\
\label{eq:egcl_lastlast}     h^{l+1}_{i} &= \phi_{h} (h^l_i, m_i, n_i)
\end{align}

Equations \ref{eq:egcl} and \ref{eq:egcl_agg} are the same message passing and aggregation over edges as in EGNNs. We add additional missing edge features $n_{ij}$, which are updated in a similar fashion with a different neural network $\phi_f$ and receive the previous timestep's missing edge embeddings if self conditioning is applied (Equations \ref{eq:self_cond} and \ref{eq:e_if}). Node positions are updated using a sum over all relative distances $(x^l_i - x^l_j)$ \citep{satorras2022enequivariantgraphneural} multiplied by the sum of the outputs of $\phi_x$ and $\phi_y$, which take in the known edge embeddings $m$ and missing edge embeddings $n$, respectively, and output scalar values (Equation \ref{eq:egcl_last}). Both edge features are then aggregated across all neighbors of each node $\mathcal{N}(i)$ (Equations \ref{eq:egcl_agg} and \ref{eq:eif_agg}) and passed to a node MLP that updates node features (Equation \ref{eq:egcl_lastlast}). We compute an MSE loss on $x$ and a Binary Cross Entropy Loss on $n_{ij}$.

As shown in the above equations, candidate edge embeddings influence atom positions (Equation \ref{eq:egcl_last}), and atom positions affect candidate edge embeddings at the following denoising step (Equation \ref{eq:e_if}). This applies to both ligand and pocket atoms --- even though pocket atoms are treated as static, the conditioning on the protein pocket guides the movement of atoms and the bond generation possibilities.

\subsection{Inference}
During inference, we follow the reverse diffusion process as described in \cite{hoogeboom2022equivariantdiffusionmoleculegeneration}, with added self conditioning \citep{chen2023analogbitsgeneratingdiscrete} using the previous timestep's node positions and  edge weights. However, we ensure that fragments stay rigid during each step of denoising using the Kabsch algorithm (Equation \ref{eq:kabsch}) \citep{lawrence2019purely} to calculate the optimal rigid body transformation:

\begin{align}
    \label{eq:kabsch} \min_{T^*_t, R^*_t} \mathcal{L}(T_t, R_t) &= \frac{1}{2} \sum_{i\in \mathcal{F}} || \hat{x}_{i,t+1} - R(\hat{x}_{i,t} + T) ||^2 \\
    \label{eq:inference_udpate} x_{t+1} &= T^*_t + R^*_t \hat{x}_{t}
\end{align}

$\hat{x}_t$ is the predicted locations of atoms in a fragment $\mathcal{F}$ at timestep $t$. $R^*$ and $T^*$ are the optimal rotational and translation vectors, respectively.

After denoising, we obtain our final atom positions and weights for every potential inter-fragment bond within $N$ Ångstroms ($N$ is a cutoff specified as a hyperparameter). To obtain our final list of bonds, we sequentially pop off the bond with the highest weight, check if the bond is chemically plausible and the associated fragments are not connected, and connect the atoms. Full details can be found in Algorithm \ref{alg:generate} in the Appendix.

\section{Results}

\subsection{Dataset \& Setup}

We follow the dataset preparation steps outlined in \cite{powers2023geometric} and \cite{medsage}. Our dataset comprises approximately 35,000 protein-ligand complexes from the Protein Data Bank (PDB). We filter out lipids, peptides, nucleic acids, and carbohydrates, and small molecules that are not considered drug-like. The resulting dataset consists of experimentally determined structures of proteins bound to high-affinity, synthesizable, drug-like small molecules.  We then split this dataset into train, validation, and test sets (70/15/15), ensuring that the proteins in any given set have less than 30\% sequence similarity to any protein in the other sets. 

To define the molecule assembly task on this dataset, we take each ligand (i.e., each small molecule) and decompose it into fragments, following the procedure and fragment library described in \cite{powers2023geometric} and \cite{medsage}. Our library comprises fragments that are small enough to be treated as rigid, such as phenyl, methyl, ethyl groups, and benzene rings. Some fragments in our dataset include only one non-hydrogen atom, whereas others include an entire aromatic ring system.

We then assign each fragment a random, independent orientation (with all orientations equally likely) by rotating around its center of mass. Finally, we translate each fragment by an random, independent 3D vector chosen from a Gaussian distribution. 

\subsection{Metrics}
To evaluate model performance, we report the following five metrics. The first four metrics quantify a model's effectiveness at predicting bonds between fragments---the main goal of molecule assembly. The fifth metric quantifies the extent to which atom positions generated by a model match those in the experimentally determined structure.

\paragraph{Precision \& Recall}
We define precision and recall as follows:

\begin{align}
    \label{eq:precision} \text{Precision} &= \frac{[\text{Predicted Bonds}] \cap \text{[True Bonds]}}{\text{Predicted Bonds}}\\
    \label{eq:recall} \text{Recall} &= \frac{[\text{Predicted Bonds}] \cap \text{[True Bonds]}}{\text{True Bonds}}
\end{align}

\paragraph{Full Molecule Recovery (FMR)}
We define ``Full Molecule Recovery'' as a binary value for each molecule: 0 if the recall is less than 1, and 1 otherwise.

\paragraph{Tanimoto Similarity}
We calculate the Tanimoto coefficient of our recapitulated molecule and the true molecule by first constructing a Morgan fingerprint \citep{rogers2010extended} of both the predicted and original molecule.  We use RDKit \citep{landrum2013rdkit} to generate the Morgan fingerprint, and use RDKit's builtin Tanimoto Similarity function to calculate the Tanimoto coefficient.

\paragraph{Root Mean Square Deviation (RMSD)}

We also calculate Root Mean Square Deviation (RMSD)---the L2 error between the predicted atom positions and atom positions in the experimentally determined structure (Equation \ref{eq:RMSD}). For the molecule assembly task, predicted atom positions are much less important than predicted bonds, but we include this metric because the results may still be instructive.

\begin{align}
    \label{eq:RMSD} \text{RMSD}(w, v) &= \sqrt{\frac{1}{n} \sum^n_{i=1} [(v_{ix} - w_{ix})^2 + (v_{iy} - w_{iy})^2 + (v_{iz} - w_{iz})^2]}
\end{align}

\subsection{Experimental Setup}

We split our comparisons table into three types of methods: EdGr, geometry-prediction models (EDM \citep{hoogeboom2022equivariantdiffusionmoleculegeneration} and Equiformer \citep{liao2022equiformer}), and topology-prediction models (Graph Convolutions \citep{kipf2016semi}, Graph Attention \citep{velivckovic2017graph}, Minimum Distance heuristic, and Common Neighbors heuristic \citep{newman2001clustering}). We train the geometry-prediction models to denoise the 3D coordinates in an attempt to recover original atom positions, and then connect the two closest atoms belonging to distinct fragments. For the topology-prediction models, we pass in the molecular graph, treating relative positions between the atom coordinates as edge features, and run standard edge prediction. We do not report RMSD for topology-prediction methods, as they do not change spatial coordinates.

\begin{figure}
    \centering
    \includegraphics[width=1.0\linewidth]{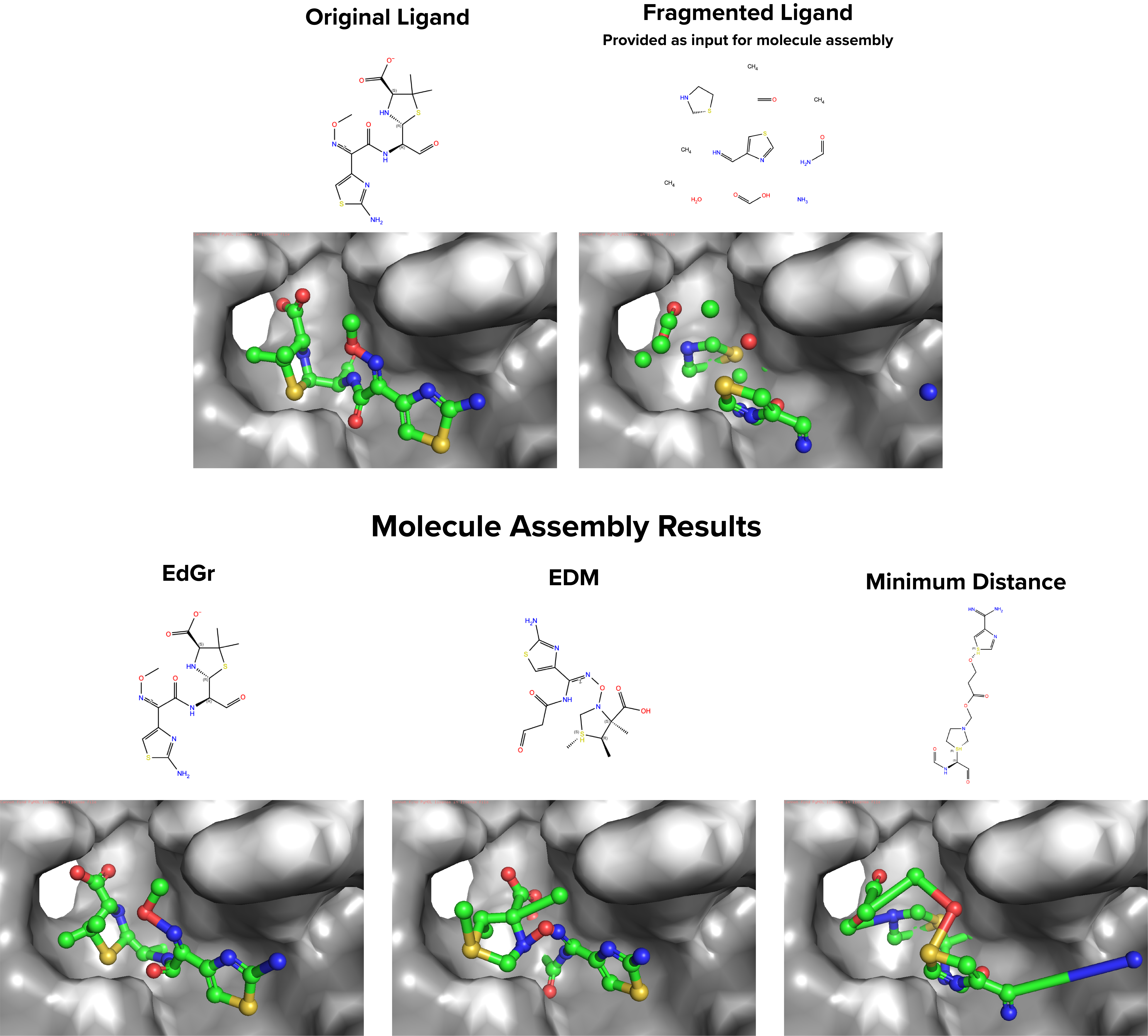}
    \caption{Examples of molecule assembly results from different methods, showing both a 2D graph depiction and a 3D rendering of each molecule. From top to bottom, left to right: the original ligand (a modified version of penicillin, from PDB entry 1LLB); the ligand decomposed into fragments, with rotational and translational noise added to each fragment; molecule assembly results from EdGr, EDM, and the Minimum Distance heuristic. }
    \label{fig:model_comparisons}
\end{figure}

\begin{table}[ht]
    \caption{Comparison of EdGr to other molecule assembly methods, with translational noise of 1Å standard deviation. Here and in the subsequent tables below, rotational noise is distributed uniformly on $SO(3)$, and error bars show 95 percent confidence intervals determined using bootstrapping. RMSD values are not listed for topology prediction methods because those methods do not adjust atom positions.}
    \label{comparisons1}
    \centering
    \begin{tabular}{c|cccc|c}
        \toprule
        \multicolumn{1}{c|}{} & \multicolumn{4}{c|}{Topology} & \multicolumn{1}{c}{Geometry} \\[0.50ex]
        Model & Precision $\uparrow$ & Recall $\uparrow$ & FMR $\uparrow$ & Tanimoto $\uparrow$ & RMSD $\downarrow$ \\
        \midrule
        EdGr & \textbf{85 $\pm$ 1\%} & \textbf{86 $\pm$ 1\%} & \textbf{64 $\pm$ 2\%} & \textbf{88 $\pm$ 1\%} & \textbf{1.09 $\pm$ 0.02Å} \\
        \midrule
        EDM & 70 $\pm$ 1\% & 70 $\pm$ 1\% & 38 $\pm$ 2\% & 71 $\pm$ 1\% & 1.20 $\pm$ 0.02Å \\
        Equiformer & 10 $\pm$ 1\% & 11 $\pm$ 1\% & 1 $\pm$ 0\% & 22 $\pm$ 0\% & 4.46 $\pm$ 0.03Å \\
        \midrule
        GCN & 23 $\pm$ 1\% & 21 $\pm$ 1\% & 2 $\pm$ 0\% & 29 $\pm$ 0\% & --- \\
        Graph Attention & 7 $\pm$ 1\% & 7 $\pm$ 1\% & 1 $\pm$ 0\% & 21 $\pm$ 0\% & --- \\
        Minimum Distance & 27 $\pm$ 1\% & 27 $\pm$ 1\% & 2 $\pm$ 1\% & 31 $\pm$ 0\% & --- \\
        Common Neighbors & 10 $\pm$ 1\% & 10 $\pm$ 1\% & 1 $\pm$ 0\% & 22 $\pm$ 0\% & --- \\
        \bottomrule
    \end{tabular}
\end{table}

\begin{table}[ht]
    \caption{Comparison of EdGr to to other molecule assembly methods, with translational noise of 2Å standard deviation.}
    \label{comparisons2}
    \centering
    \begin{tabular}{c|cccc|c}
        \toprule
        \multicolumn{1}{c|}{} & \multicolumn{4}{c|}{Topology} & \multicolumn{1}{c}{Geometry} \\[0.50ex]
        Model & Precision $\uparrow$ & Recall $\uparrow$ & FMR $\uparrow$ & Tanimoto $\uparrow$ & RMSD $\downarrow$ \\
        \midrule
        EdGr & \textbf{76 $\pm$ 1\%} & \textbf{77 $\pm$ 1\%} & \textbf{44 $\pm$ 2\%} & \textbf{74 $\pm$ 1\%} & \textbf{1.58 $\pm$ 0.02Å} \\
        \midrule
        EDM & 59 $\pm$ 1\% & 60 $\pm$ 1\% & 24 $\pm$ 1\% & 61 $\pm$ 1\% & 1.65 $\pm$ 0.03Å \\
        Equiformer & 10 $\pm$ 1\% & 10 $\pm$ 1\% & 1 $\pm$ 0\% & 22 $\pm$ 0\% & 5.25 $\pm$ 0.04Å \\
        \midrule
        GCN & 11$\pm$ 1\% & 11 $\pm$ 1\% & 1 $\pm$ 0\% & 22 $\pm$ 0\% & --- \\
        Graph Attention & 7 $\pm$ 1\% & 7 $\pm$ 0\% & 1 $\pm$ 0\% & 21 $\pm$ 0\% & --- \\
        Minimum Distance & 19 $\pm$ 1\% & 19 $\pm$ 1\% & 1 $\pm$ 0\% & 26 $\pm$ 0\% & --- \\
        Common Neighbors & 8 $\pm$ 1\% & 8 $\pm$ 1\% & 1 $\pm$ 0\% & 21 $\pm$ 0\% & --- \\
        \bottomrule
    \end{tabular}
\end{table}

\begin{table}[ht]
    \caption{Comparison of EdGr to to other molecule assembly methods, with translational noise of 3Å standard deviation.}
    \label{comparisons3}
    \centering
    \begin{tabular}{c|cccc|c}
        \toprule
        \multicolumn{1}{c|}{} & \multicolumn{4}{c|}{Topology} & \multicolumn{1}{c}{Geometry} \\[0.50ex]
        Model & Precision $\uparrow$ & Recall $\uparrow$ & FMR $\uparrow$ & Tanimoto $\uparrow$ & RMSD $\downarrow$ \\
        \midrule
        EdGr & \textbf{71 $\pm$ 1\%} & \textbf{70 $\pm$ 1\%} & \textbf{34 $\pm$ 2\%} & \textbf{72 $\pm$ 1\%} & \textbf{1.96 $\pm$ 0.03Å} \\
        \midrule
        EDM & 52 $\pm$ 1\% & 52 $\pm$ 2\% & 15 $\pm$ 1\% & 51 $\pm$ 1\% & 2.00 $\pm$ 0.03Å \\
        Equiformer & 10 $\pm$ 1\% & 11 $\pm$ 1\% & 1 $\pm$ 0\% & 22 $\pm$ 0\% & 6.43 $\pm$ 0.04Å \\
        \midrule
        GCN & 10 $\pm$ 1\% & 10 $\pm$ 1\% & 1 $\pm$ 0\% & 21 $\pm$ 0\% & --- \\
        Graph Attention & 8 $\pm$ 1\% & 7 $\pm$ 1\% & 1 $\pm$ 0\% & 23 $\pm$ 0\% & --- \\
        Minimum Distance & 15 $\pm$ 1\% & 15 $\pm$ 1\% & 1 $\pm$ 0\% & 24 $\pm$ 0\% & --- \\
        Common Neighbors & 8 $\pm$ 1\% & 8 $\pm$ 1\% & 1 $\pm$ 0\% & 21 $\pm$ 0\% & --- \\
        \bottomrule
    \end{tabular}
\end{table}

\subsection{Comparisons}

To evaluate model robustness, we report model performance on different amounts of noise added to the fragments. We report differing amounts of \textit{translational} noise $\mathcal{N}(\mu, \sigma^2)$, where we test $\sigma=1$Å (Table \ref{comparisons1}), $\sigma=2$Å (Table \ref{comparisons2}), and $\sigma=3$Å (Table \ref{comparisons3}). A fragment's \textit{rotational} noise is always sampled uniformly from $SO(3)$, meaning that all rotations are equally likely. 

To generate confidence intervals, we generate three samples for each ligand in our test dataset (roughly 1,500 examples). We then perform bootstrap sampling to generate a 95 percent confidence interval.

EdGr outperforms all other models tested according to every metric at every level of translational noise. The next-best method is EDM, which makes uses diffusion to iteratively refine atom positions over $N$ diffusion timesteps. Equiformer attempts to predict the final denoised position in a one-shot fashion, and does poorly. The topology-prediction methods performed poorly across the board, likely due to their inability to refine coordinate positions, leading to incorrect bond predictions. We see this trend as the noise levels increase --- GCN and Minimum Distance, the best-performing topology-prediction methods, deteriorate in performance. However, EdGr exhibits robust performance despite the increasing amount of noise, with only a 14\% drop in precision and a 0.87Å increase in RMSD as the translational noise level increases from $\sigma=1$Å to $\sigma=3$Å.


\subsection{Ablation Study}

In Table \ref{ablations}, we report an ablation study. Removing self conditioning yielded a drop in performance. This was expected, as knowing the model's confidence in the predicted bonds at the previous timestep of denoising should yield an improvement in the following denoising timestep's predictions.

\section{Conclusion}
We present EdGr, a graph diffusion-based edge prediction method for molecule assembly that couples prediction of additional bonds with adjustment of atom positions. EdGr substantially and consistently outperforms previous methods for this task, which is important in drug design.

EdGr has several limitations in its current form, as discussed in the Appendix. We believe these can be addressed in future work.

The innovations underlying EdGr---explicit supervision of edge likelihoods and coupled diffusion over coordinates and connectivity---offer a general framework for spatial graph completion. This framework is applicable, in principle, to any graph completion task in which nodes have spatial coordinates that influence edge likelihood and in which knowledge of missing edges would help determine spatial coordinates. For example, in neural connectomics, one wishes to infer fully connected neural circuits from microscopy data in which many connections between neurons are not visible and precise geometries of neurons are uncertain \citep{ding2025functional, marc2013retinal}. In the computer vision problem of 3D scene reconstruction, one wishes both to determine relationships between objects and to correct for spatial misalignments between objects in images from different view angles \citep{koch2024sgrec3d}.  When designing a wireless sensor network, one must determine both spatial positions of sensors and connectivity between sensors \citep{khojasteh2022node, dogan2017uncertainty}. Our results may thus have implications well beyond molecular design.

\begin{table}
  \caption{Ablation study of EdGr, with translational noise of 1Å standard deviation.}
  \label{ablations}
  \centering
  \begin{tabular}{c|cccc|c}
    \toprule
    \multicolumn{1}{c|}{} & \multicolumn{4}{c|}{Topology} & \multicolumn{1}{c}{Geometry} \\[0.5ex]
    Model & Precision $\uparrow$ & Recall $\uparrow$ & FMR $\uparrow$ & Tanimoto $\uparrow$ & RMSD $\downarrow$ \\
    \midrule
    Base model & 85 $\pm$ 1\% & 86 $\pm$ 1\% & 64 $\pm$ 2\% & 88 $\pm$ 1\% & 1.09 $\pm$ 0.02Å \\
    No self conditioning & 80 $\pm$ 1\% & 79 $\pm$ 1\% & 52 $\pm$ 2\% & 83 $\pm$ 1\% & 1.28 $\pm$ 0.02Å \\
    \bottomrule
  \end{tabular}
\end{table}

\appendix

\section{Technical Appendices and Supplementary Material}

\subsection{Broader Societal Impacts} We believe that this work is important to the field of structure-based drug design, which is highly useful for creating novel drugs to treat diseases and improve human health. However, any such method could also potentially be used to create drugs that do harm. Care is necessary to ensure that this method is used for beneficial purposes.

\subsection{Limitations} In our current setup, we treat fragments as rigid. This is a reasonable assumption for the fragments in our current library, which do not contain any rotatable bonds. To use EdGr with fragments that do contain rotatable bonds, one would need to retrain it to allow rotation around these bonds.

In addition, the current version of EdGr only predicts single bonds---as opposed to double or triple bonds---between fragments. EdGr could be extended to predict different bond types as well. This would again require retraining.

\subsection{Final Bond Selection Algorithm}

\begin{algorithm}
\caption{Final Bond Selection}
\label{alg:generate}
\begin{algorithmic}[1]
\Require List of bonds and model weights for each, ordered from lowest to highest: \texttt{bonds}
\State Initialize QuickFind datatype: \texttt{q(n={num\_fragments})}
\While{\texttt{!q.is\_fully\_connected()}}
    \State \texttt{atom1, atom2 = bonds.pop()}
    \State \texttt{f1, f2 = get\_fragment(atom1), get\_fragment(atom2)}
    \If{\texttt{atom1} and \texttt{atom2} are bonded to hydrogen atoms and \texttt{!q.is\_connected(f1,f2)}}
        \State Add \texttt{atom1} and \texttt{atom2} to final list of bonds
        \State \texttt{q.connect(f1,f2)}
    \Else
        \State \texttt{continue}
    \EndIf
\EndWhile
\State \Return final list of bonds
\end{algorithmic}
\end{algorithm}

\subsection{Training \& Reproducibility Details for EdGr}
We train our models on a single Nvidia GPU for up to 300 epochs (approximately 1 week on an Nvidida A40), using the checkpoint with the lowest validation loss for benchmarking. We train all our diffusion models with the AdamW Optimizer, with a learning rate of $3 \times 10^{-4}$, with 100 diffusion steps, batch normalization, using ReLU activations, with 4 hidden layers, each comprising 128 neurons. EdGr receives atom coordinates, element types encoded as one-hot vectors, and fragment membership encoded as a binary vector as input features.

\subsection{Training \& Reproducibility Details for Geometry-Prediction}

EDM and Equiformer receive the same input features as EdGr.

\paragraph{EDM} We use the EDM architecture from the DiffLinker \citep{igashov2024equivariant} codebase. We use the same hyperparameters as EdGr ($3 \times 10^{-4}$ learning rate, 4 EGCL layers, each comprising 128 neurons, AdamW optimizer, ReLU activations, 100 diffusion steps, batch normalization). We treat pocket atoms as static and all ligand atoms as flexible. We trained EDM for 300 epochs, saving the checkpoint with the lowest validation loss and using that for benchmarking. To generate the final inter-fragment bonds, we use Algorithm \ref{alg:generate}, but instead of using the \texttt{bonds} list ordered by model weight, the \texttt{bonds} list is ordered by Euclidean distance.

\paragraph{Equiformer} We use the implementation of Equiformer \citep{liao2022equiformer} at this GitHub repository: https://github.com/lucidrains/equiformer-pytorch. Due to compute constraints---training an Equiformer model on a single A100 GPU took over a week, with 1 epoch completing every 2 hours---we could not train Equiformer for the full 300 epochs and instead trained it for a week on an A100 (roughly 80 epochs). We saved the model with the lowest validation loss and used that checkpoint for benchmarking. We used the default hyperparameters from the repository, but modified the following: \texttt{num\_edge\_tokens = 2, edge\_dim = 4, single\_headed\_kv = True, heads = 4, dim\_head = 8}. We generate inter-fragment bonds in the same manner as described in the paragraph describing running EDM.

\subsection{Training \& Reproducibility Details for Topology-Prediction}

As additional baselines, we also tested standard implementations of the Graph Convolution Network (GCN) \citep{kipf2016semi} and Graph Attention Networks \citep{velivckovic2017graph} from PyTorch Geometric Version 2.7 (https://pytorch-geometric.readthedocs.io/en/latest/index.html). For both architectures we create a node embedding model that learned node embeddings based on the molecular graphs of all of the provided fragments and a separate link prediction network that took pairs of node embeddings and predicted whether they formed an inter-fragment bond. Each model had nodes that represented atoms, with node features including a one-hot representation of element type, a numerical representation of the fragment identity, the atom's current valence, the maximum number of bonds the atom could form, and a binary flag of whether the atom could form any additional bonds. Learning rates of $1 \times 10^{-2}$, $1 \times 10^{-3}$, and $1 \times 10^{-4}$ were tested for both models. Both methods were trained for 3 epochs on a single GPU with the default settings for the AdamW optimizer (https://arxiv.org/abs/1711.05101), with the model checkpoints at the end of each epoch featuring the lowest validation loss across all hyperparameters being used to report metrics. During development, additional hyperparameter settings beyond those listed below and longer training times including up to 20 epochs were tested, but did not result in significant changes in validation loss or validation performance. The results reported correspond to the best results obtained from the combination of hyperparameters investigated for these models.

\paragraph{Running GCN} The GCN node embedding network featured $3$ GCN layers, with a hidden dimension size of $128$, and a node embedding output dimension of size $64$. The inverted pairwise distances between all atoms based on the noised $3$D coordinates were used as edge weights for message passing in the GCN. The ReLU function was used as the non-linear activation function, and dropout layers were placed after the ReLU activation for the first two GCN layers with a dropout rate of $10\%$. The link prediction network featured $3$ MLP layers, with ReLU as the activation function and dropout layers after the first two MLP layers with a dropout rate of $10\%$. The input dimension for the link prediction network was $128$, equal to a pair of node embeddings concatenated together, and the hidden dimension was size $64$. The output dimension was size $1$, for the binary classification task of whether a given pair of node embeddings should have an inter-fragment bond.

\paragraph{Running Graph Attention} The Graph Attention node embedding network featured $3$ Graph Attention layers, with a hidden dimension size of $128$, a node embedding output dimension of size $64$, and $4$ attention heads. The pairwise distances between all atoms based on the noised $3$D coordinates were provided as edge attributes to each node, along with a one-hot encoding representing whether a given edge was a known intra-fragment bond or a potential inter-fragment bond. The ReLU function was used as the non-linear activation function, and dropout layers were placed after the ReLU activation for the first two GCN layers with a dropout rate of $30\%$. The link prediction network featured $3$ MLP layers, with ReLU as the activation function and dropout layers after the first two MLP layers with a dropout rate of $30\%$. The input dimension for the link prediction network was $128$, equal to a pair of node embeddings concatenated together. The output dimension was size $1$, for the binary classification task of whether a given pair of node embeddings should have an inter-fragment bond.

\subsection{Additional Result}

We have previously mentioned the importance of direct edge prediction and the usage of these weights to influence node positions. We investigate the importance of the latter in the following ablation study. We continue to output logits $n_{ij}$ for every candidate edge, but remove these terms from the node coordinate and feature updates. To be precise, we remove the $\phi_y(n_{ij})$ from Equation \ref{eq:egcl_last} and $n_{i}$ from Equation \ref{eq:egcl_lastlast} to obtain the following:

\begin{align}
\label{eq:no_coupling_egcl} x^{l+1}_{i} &= x^{l}_{i} + \frac{1}{M-1} \sum_{j \ne i} (x^l_i - x^l_j) \phi_x(m_{ij}) \\
\label{eq:no_coupling_gcl} h^{l+1}_{i} &= \phi_{h} (h^l_i, m_i)
\end{align}

We note that these ablated formulas (Equations \ref{eq:no_coupling_egcl} and \ref{eq:no_coupling_gcl}) are identical to the coordinate and node feature update formulas in standard EGCL layers used in EGNNs \citep{satorras2022enequivariantgraphneural}. In other words, one could say that we are evaluating the performance of direct edge prediction while still using the original EGNN coordinate and feature updates.

We call this ablation ``Remove candidate edge$\rightarrow$node update," and the results can be found in Table \ref{additional}. We find that removing this update while maintaining direct edge prediction results in significantly reduced performance, highlighting the importance of using the candidate edge weights $n_{ij}$ in the updates to the node features $h^l_i$ and coordinates $x_i^l$.

\begin{table}
  \caption{Ablation study of EdGr, with translational noise of 1Å standard deviation.}
  \label{additional}
  \centering
  \begin{tabular}{c|cccc|c}
    \toprule
    \multicolumn{1}{c|}{} & \multicolumn{4}{c|}{Topology} & \multicolumn{1}{c}{Geometry} \\[0.5ex]
    Model & Precision $\uparrow$ & Recall $\uparrow$ & FMR $\uparrow$ & Tanimoto $\uparrow$ & RMSD $\downarrow$ \\
    \midrule
    EdGr base model & 85 $\pm$ 1\% & 86 $\pm$ 1\% & 64 $\pm$ 2\% & 88 $\pm$ 1\% & 1.09 $\pm$ 0.02Å \\[0.75ex]
    \makecell[l]{Remove candidate \\ edge$\rightarrow$node update} & 63 $\pm$ 1\% & 64 $\pm$ 1\% & 26 $\pm$ 1\% & 63 $\pm$ 1\% & 1.33 $\pm$ 0.02Å \\
    \bottomrule
  \end{tabular}
\end{table}

\clearpage
\bibliographystyle{plainnat}
\bibliography{neurips_2025_preprint}

\begin{thebibliography}{40}
\providecommand{\natexlab}[1]{#1}
\providecommand{\url}[1]{\texttt{#1}}
\expandafter\ifx\csname urlstyle\endcsname\relax
  \providecommand{\doi}[1]{doi: #1}\else
  \providecommand{\doi}{doi: \begingroup \urlstyle{rm}\Url}\fi

\bibitem[Carloni et~al.(2025)Carloni, Rossetti, and M\"uller]{carloni2025rational}
Paolo Carloni, Giulia Rossetti, and Christa~E M\"uller.
\newblock Rational design of ligands with optimized residence time.
\newblock \emph{ACS Pharmacology \& Translational Science}, 2025.

\bibitem[Chaudhri et~al.(2021)Chaudhri, Chittar, and Genesereth]{chaudhri2021introduction}
Vinay~K. Chaudhri, Naren Chittar, and Michael Genesereth.
\newblock An introduction to knowledge graphs.
\newblock \url{https://ai.stanford.edu/blog/introduction-to-knowledge-graphs/}, May 2021.
\newblock Stanford AI Lab Blog.

\bibitem[Chen et~al.(2023)Chen, Zhang, and Hinton]{chen2023analogbitsgeneratingdiscrete}
Ting Chen, Ruixiang Zhang, and Geoffrey Hinton.
\newblock Analog bits: Generating discrete data using diffusion models with self-conditioning, 2023.
\newblock URL \url{https://arxiv.org/abs/2208.04202}.

\bibitem[Ding et~al.(2025)Ding, Fahey, Papadopoulos, Wang, Celii, Papadopoulos, Chang, Kunin, Tran, Fu, et~al.]{ding2025functional}
Zhuokun Ding, Paul~G Fahey, Stelios Papadopoulos, Eric~Y Wang, Brendan Celii, Christos Papadopoulos, Andersen Chang, Alexander~B Kunin, Dat Tran, Jiakun Fu, et~al.
\newblock Functional connectomics reveals general wiring rule in mouse visual cortex.
\newblock \emph{Nature}, 640\penalty0 (8058):\penalty0 459--469, 2025.

\bibitem[Dogan and Brown(2017)]{dogan2017uncertainty}
Gulustan Dogan and Ted Brown.
\newblock Uncertainty modeling in wireless sensor networks.
\newblock In \emph{Proceedings of the International Conference on Big Data and Internet of Thing}, BDIOT '17, page 200–204, New York, NY, USA, 2017. Association for Computing Machinery.
\newblock ISBN 9781450354301.
\newblock \doi{10.1145/3175684.3175692}.
\newblock URL \url{https://doi.org/10.1145/3175684.3175692}.

\bibitem[Du et~al.(2024)Du, Jamasb, Guo, Fu, Harris, Wang, Duan, Li{\`o}, Schwaller, and Blundell]{du2024machine}
Yuanqi Du, Arian~R Jamasb, Jeff Guo, Tianfan Fu, Charles Harris, Yingheng Wang, Chenru Duan, Pietro Li{\`o}, Philippe Schwaller, and Tom~L Blundell.
\newblock Machine learning-aided generative molecular design.
\newblock \emph{Nature Machine Intelligence}, 6\penalty0 (6):\penalty0 589--604, 2024.

\bibitem[Guan et~al.(2023)Guan, Qian, Peng, Su, Peng, and Ma]{guan20233d}
Jiaqi Guan, Wesley~Wei Qian, Xingang Peng, Yufeng Su, Jian Peng, and Jianzhu Ma.
\newblock 3d equivariant diffusion for target-aware molecule generation and affinity prediction.
\newblock \emph{arXiv preprint arXiv:2303.03543}, 2023.

\bibitem[Guan et~al.(2024)Guan, Zhou, Yang, Bao, Peng, Ma, Liu, Wang, and Gu]{guan2024decompdiff}
Jiaqi Guan, Xiangxin Zhou, Yuwei Yang, Yu~Bao, Jian Peng, Jianzhu Ma, Qiang Liu, Liang Wang, and Quanquan Gu.
\newblock Decompdiff: diffusion models with decomposed priors for structure-based drug design.
\newblock \emph{arXiv preprint arXiv:2403.07902}, 2024.

\bibitem[Ho et~al.(2020)Ho, Jain, and Abbeel]{ho2020denoisingdiffusionprobabilisticmodels}
Jonathan Ho, Ajay Jain, and Pieter Abbeel.
\newblock Denoising diffusion probabilistic models, 2020.
\newblock URL \url{https://arxiv.org/abs/2006.11239}.

\bibitem[Hoogeboom et~al.(2022)Hoogeboom, Satorras, Vignac, and Welling]{hoogeboom2022equivariantdiffusionmoleculegeneration}
Emiel Hoogeboom, Victor~Garcia Satorras, Clément Vignac, and Max Welling.
\newblock Equivariant diffusion for molecule generation in 3d, 2022.
\newblock URL \url{https://arxiv.org/abs/2203.17003}.

\bibitem[Huang et~al.(2022)Huang, Peng, Ma, and Zhang]{huang20223dlinker}
Yinan Huang, Xingang Peng, Jianzhu Ma, and Muhan Zhang.
\newblock 3dlinker: an e (3) equivariant variational autoencoder for molecular linker design.
\newblock \emph{arXiv preprint arXiv:2205.07309}, 2022.

\bibitem[Igashov et~al.(2024)Igashov, St{\"a}rk, Vignac, Schneuing, Satorras, Frossard, Welling, Bronstein, and Correia]{igashov2024equivariant}
Ilia Igashov, Hannes St{\"a}rk, Cl{\'e}ment Vignac, Arne Schneuing, Victor~Garcia Satorras, Pascal Frossard, Max Welling, Michael Bronstein, and Bruno Correia.
\newblock Equivariant 3d-conditional diffusion model for molecular linker design.
\newblock \emph{Nature Machine Intelligence}, 6\penalty0 (4):\penalty0 417--427, 2024.

\bibitem[Imrie et~al.(2020)Imrie, Bradley, van~der Schaar, and Deane]{imrie2020deep}
Fergus Imrie, Anthony~R Bradley, Mihaela van~der Schaar, and Charlotte~M Deane.
\newblock Deep generative models for 3d linker design.
\newblock \emph{Journal of chemical information and modeling}, 60\penalty0 (4):\penalty0 1983--1995, 2020.

\bibitem[Khojasteh et~al.(2022)Khojasteh, Saucan, Liu, Conti, and Win]{khojasteh2022node}
Mohammad~Javad Khojasteh, Augustin~A Saucan, Zhenyu Liu, Andrea Conti, and Moe~Z Win.
\newblock Node deployment under position uncertainty for network localization.
\newblock In \emph{ICC 2022-IEEE International Conference on Communications}, pages 889--894. IEEE, 2022.

\bibitem[Kipf and Welling(2016)]{kipf2016semi}
Thomas~N Kipf and Max Welling.
\newblock Semi-supervised classification with graph convolutional networks.
\newblock \emph{arXiv preprint arXiv:1609.02907}, 2016.

\bibitem[Koch et~al.(2024)Koch, Hermosilla, Vaskevicius, Colosi, and Ropinski]{koch2024sgrec3d}
Sebastian Koch, Pedro Hermosilla, Narunas Vaskevicius, Mirco Colosi, and Timo Ropinski.
\newblock Sgrec3d: Self-supervised 3d scene graph learning via object-level scene reconstruction.
\newblock In \emph{Proceedings of the IEEE/CVF Winter Conference on Applications of Computer Vision}, pages 3404--3414, 2024.

\bibitem[Lamoree and Hubbard(2017)]{lamoree2017current}
Bas Lamoree and Roderick~E Hubbard.
\newblock Current perspectives in fragment-based lead discovery (fbld).
\newblock \emph{Essays in biochemistry}, 61\penalty0 (5):\penalty0 453--464, 2017.

\bibitem[Landrum(2013)]{landrum2013rdkit}
Greg Landrum.
\newblock Rdkit documentation.
\newblock \emph{Release}, 1\penalty0 (1-79):\penalty0 4, 2013.

\bibitem[Lawrence et~al.(2019)Lawrence, Bernal, and Witzgall]{lawrence2019purely}
Jim Lawrence, Javier Bernal, and Christoph Witzgall.
\newblock A purely algebraic justification of the kabsch-umeyama algorithm.
\newblock \emph{Journal of research of the National Institute of Standards and Technology}, 124:\penalty0 1, 2019.

\bibitem[Leach et~al.(2022)Leach, Schmon, Degiacomi, and Willcocks]{leach2022denoising}
Adam Leach, Sebastian~M Schmon, Matteo~T. Degiacomi, and Chris~G. Willcocks.
\newblock Denoising diffusion probabilistic models on {SO}(3) for rotational alignment.
\newblock In \emph{ICLR 2022 Workshop on Geometrical and Topological Representation Learning}, 2022.
\newblock URL \url{https://openreview.net/forum?id=BY88eBbkpe5}.

\bibitem[Li et~al.(2023)Li, Shomer, Mao, Zeng, Ma, Shah, Tang, and Yin]{li2023evaluating}
Juanhui Li, Harry Shomer, Haitao Mao, Shenglai Zeng, Yao Ma, Neil Shah, Jiliang Tang, and Dawei Yin.
\newblock Evaluating graph neural networks for link prediction: Current pitfalls and new benchmarking.
\newblock \emph{Advances in Neural Information Processing Systems}, 36:\penalty0 3853--3866, 2023.

\bibitem[Liao and Smidt(2022)]{liao2022equiformer}
Yi-Lun Liao and Tess Smidt.
\newblock Equiformer: Equivariant graph attention transformer for 3d atomistic graphs.
\newblock \emph{arXiv preprint arXiv:2206.11990}, 2022.

\bibitem[Mao et~al.(2023)Mao, Li, Shomer, Li, Fan, Ma, Zhao, Shah, and Tang]{mao2023revisiting}
Haitao Mao, Juanhui Li, Harry Shomer, Bingheng Li, Wenqi Fan, Yao Ma, Tong Zhao, Neil Shah, and Jiliang Tang.
\newblock Revisiting link prediction: A data perspective.
\newblock \emph{arXiv preprint arXiv:2310.00793}, 2023.

\bibitem[Marc et~al.(2013)Marc, Jones, Watt, Anderson, Sigulinsky, and Lauritzen]{marc2013retinal}
Robert~E Marc, Bryan~W Jones, Carl~B Watt, James~R Anderson, Crystal Sigulinsky, and Scott Lauritzen.
\newblock Retinal connectomics: towards complete, accurate networks.
\newblock \emph{Progress in retinal and eye research}, 37:\penalty0 141--162, 2013.

\bibitem[Morehead and Cheng(2024)]{morehead2024geometry}
Alex Morehead and Jianlin Cheng.
\newblock Geometry-complete diffusion for 3d molecule generation and optimization.
\newblock \emph{Communications Chemistry}, 7\penalty0 (1):\penalty0 150, 2024.

\bibitem[Neeser et~al.(2025)Neeser, Igashov, Schneuing, Bronstein, Schwaller, and Correia]{neeser2025flowbased}
Rebecca~Manuela Neeser, Ilia Igashov, Arne Schneuing, Michael~M. Bronstein, Philippe Schwaller, and Bruno Correia.
\newblock Flow-based fragment identification via contrastive learning of binding site-specific latent representations.
\newblock In \emph{AI for Accelerated Materials Design - ICLR 2025}, 2025.
\newblock URL \url{https://openreview.net/forum?id=bZW1HLT1gI}.

\bibitem[Newman(2001)]{newman2001clustering}
Mark~EJ Newman.
\newblock Clustering and preferential attachment in growing networks.
\newblock \emph{Physical review E}, 64\penalty0 (2):\penalty0 025102, 2001.

\bibitem[O'Boyle et~al.(2011)O'Boyle, Banck, James, Morley, Vandermeersch, and Hutchison]{o2011open}
Noel~M O'Boyle, Michael Banck, Craig~A James, Chris Morley, Tim Vandermeersch, and Geoffrey~R Hutchison.
\newblock Open babel: An open chemical toolbox.
\newblock \emph{Journal of cheminformatics}, 3:\penalty0 1--14, 2011.

\bibitem[Powers et~al.(2023)Powers, Yu, Suriana, Koodli, Lu, Paggi, and Dror]{powers2023geometric}
Alexander~S Powers, Helen~H Yu, Patricia Suriana, Rohan~V Koodli, Tianyu Lu, Joseph~M Paggi, and Ron~O Dror.
\newblock Geometric deep learning for structure-based ligand design.
\newblock \emph{ACS Central Science}, 9\penalty0 (12):\penalty0 2257--2267, 2023.

\bibitem[Powers et~al.(2025)Powers, Lu, Koodli, Xu, Gu, Karelina, and Dror]{medsage}
Alexander~S Powers, Tianyu Lu, Rohan~V Koodli, Minkai Xu, Siyi Gu, Masha Karelina, and Ron~O Dror.
\newblock {MedSAGE}: {Bridging} {Generative} {AI} and {Medicinal} {Chemistry} for {Structure}-{Based} {Design} of {Small} {Molecule} {Drugs}.
\newblock \emph{bioRxiv}, 2025.
\newblock \doi{10.1101/2025.05.10.653107}.
\newblock URL \url{https://www.biorxiv.org/content/early/2025/05/15/2025.05.10.653107}.

\bibitem[Rogers and Hahn(2010)]{rogers2010extended}
David Rogers and Mathew Hahn.
\newblock Extended-connectivity fingerprints.
\newblock \emph{Journal of chemical information and modeling}, 50\penalty0 (5):\penalty0 742--754, 2010.

\bibitem[Satorras et~al.(2022)Satorras, Hoogeboom, and Welling]{satorras2022enequivariantgraphneural}
Victor~Garcia Satorras, Emiel Hoogeboom, and Max Welling.
\newblock E(n) equivariant graph neural networks, 2022.
\newblock URL \url{https://arxiv.org/abs/2102.09844}.

\bibitem[Savjolova(1985)]{savjolova}
T.~I. Savjolova.
\newblock Preface to novye metody issledovanija tekstury polikristalliceskich materialov, 1985.

\bibitem[Schneuing et~al.(2024)Schneuing, Harris, Du, Didi, Jamasb, Igashov, Du, Gomes, Blundell, Lio, et~al.]{schneuing2024structure}
Arne Schneuing, Charles Harris, Yuanqi Du, Kieran Didi, Arian Jamasb, Ilia Igashov, Weitao Du, Carla Gomes, Tom~L Blundell, Pietro Lio, et~al.
\newblock Structure-based drug design with equivariant diffusion models.
\newblock \emph{Nature Computational Science}, 4\penalty0 (12):\penalty0 899--909, 2024.

\bibitem[Schneuing et~al.(2025)Schneuing, Igashov, Dobbelstein, Castiglione, Bronstein, and Correia]{schneuing2025multidomain}
Arne Schneuing, Ilia Igashov, Adrian~W. Dobbelstein, Thomas Castiglione, Michael~M. Bronstein, and Bruno Correia.
\newblock Multi-domain distribution learning for de novo drug design.
\newblock In \emph{The Thirteenth International Conference on Learning Representations}, 2025.
\newblock URL \url{https://openreview.net/forum?id=g3VCIM94ke}.

\bibitem[Sheng and Zhang(2013)]{sheng2013fragment}
Chunquan Sheng and Wannian Zhang.
\newblock Fragment informatics and computational fragment-based drug design: an overview and update.
\newblock \emph{Medicinal Research Reviews}, 33\penalty0 (3):\penalty0 554--598, 2013.

\bibitem[Shim and MacKerell~Jr(2011)]{shim2011computational}
Jihyun Shim and Alexander~D MacKerell~Jr.
\newblock Computational ligand-based rational design: role of conformational sampling and force fields in model development.
\newblock \emph{MedChemComm}, 2\penalty0 (5):\penalty0 356--370, 2011.

\bibitem[Sohl-Dickstein et~al.(2015)Sohl-Dickstein, Weiss, Maheswaranathan, and Ganguli]{sohl2015deep}
Jascha Sohl-Dickstein, Eric Weiss, Niru Maheswaranathan, and Surya Ganguli.
\newblock Deep unsupervised learning using nonequilibrium thermodynamics.
\newblock In \emph{International conference on machine learning}, pages 2256--2265. pmlr, 2015.

\bibitem[Veli{\v{c}}kovi{\'c} et~al.(2017)Veli{\v{c}}kovi{\'c}, Cucurull, Casanova, Romero, Lio, and Bengio]{velivckovic2017graph}
Petar Veli{\v{c}}kovi{\'c}, Guillem Cucurull, Arantxa Casanova, Adriana Romero, Pietro Lio, and Yoshua Bengio.
\newblock Graph attention networks.
\newblock \emph{arXiv preprint arXiv:1710.10903}, 2017.

\bibitem[Zamini et~al.(2022)Zamini, Reza, and Rabiei]{Zamini_2022}
Mohamad Zamini, Hassan Reza, and Minou Rabiei.
\newblock A review of knowledge graph completion.
\newblock \emph{Information}, 13\penalty0 (8):\penalty0 396, August 2022.
\newblock ISSN 2078-2489.
\newblock \doi{10.3390/info13080396}.
\newblock URL \url{http://dx.doi.org/10.3390/info13080396}.

\end{thebibliography}

\end{document}